# A Platform for the Biomedical Application of Large Language Models

## Authors

- **Sebastian Lobentanzer** ✉
  [0000-0003-3399-6695](#) · [slobentanzer](#) · [slobentanzer](#)
  Heidelberg University, Faculty of Medicine and Heidelberg University Hospital, Institute for Computational Biomedicine, Heidelberg, Germany

- **Shaohong Feng**
  [0009-0000-8124-3626](#) · [fengsh27](#)
  Department of Biomedical Informatics, The Ohio State University, Columbus, Ohio, USA

- **The BioChatter Consortium**

- **Andreas Maier**
  [0000-0003-4408-0068](#) · [andimajore](#)
  Institute for Computational Systems Biology, University of Hamburg, Hamburg, Germany

- **Cankun Wang**
  [0000-0002-0225-9855](#) · [Wang-Cankun](#) · [Wang-Cankun](#)
  Department of Biomedical Informatics, The Ohio State University, Columbus, Ohio, USA

- **Jan Baumbach**
  [0000-0002-0282-0462](#)
  Institute for Computational Systems Biology, University of Hamburg, Hamburg, Germany; Computational Biomedicine Lab, Department of Mathematics and Computer Science, University of Southern Denmark, Odense, Denmark

- **Nils Krehl**
  [0009-0003-1780-6123](#) · [nilskre](#)
  Heidelberg University, Faculty of Medicine and Heidelberg University Hospital, Institute for Computational Biomedicine, Heidelberg, Germany

- **Qin Ma**
  [0000-0002-3264-8392](#) · [maqin2001](#) · [QinMaBMBL](#)
  Department of Biomedical Informatics, The Ohio State University, Columbus, Ohio, USA

- **Julio Saez-Rodriguez** ✉
  [0000-0002-8552-8976](#) · [saezrodriguez](#) · [saezlab](#)
  Heidelberg University, Faculty of Medicine and Heidelberg University Hospital, Institute for Computational Biomedicine, Heidelberg, Germany

✉ — Correspondence possible via [GitHub Issues](#) or email to Sebastian Lobentanzer <sebastian.lobentanzer@gmail.com>, Julio Saez-Rodriguez <pub.saez@uni-heidelberg.de>.

Authors between consortium and last author are ordered alphabetically.

# Abstract

Current-generation Large Language Models (LLMs) have stirred enormous interest in recent months, yielding great potential for accessibility and automation, while simultaneously posing significant challenges and risk of misuse. To facilitate interfacing with LLMs in the biomedical space, while at the same time safeguarding their functionalities through sensible constraints, we propose a dedicated, open-source framework: BioChatter. Based on open-source software packages, we synergise the many functionalities that are currently developing around LLMs, such as knowledge integration / retrieval-augmented generation, model chaining, and benchmarking, resulting in an easy-to-use and inclusive framework for application in many use cases of biomedicine. We focus on robust and user-friendly implementation, including ways to deploy privacy-preserving local open-source LLMs. We demonstrate use cases via two multi-purpose web apps (https://chat.biocypher.org), and provide documentation, support, and an open community.

# Introduction

Despite technological advances, understanding biological and biomedical systems still poses major challenges [1,2]. We measure more and more data points with ever-increasing resolution to such a degree that their analysis and interpretation have become the bottleneck for their exploitation [2]. One reason for this challenge may be the inherent limitation of human knowledge [3]: Even seasoned domain experts cannot know the implications of every gene, molecule, symptom, or biomarker. In addition, biological events are context-dependent, for instance with respect to a cell type or specific disease.

Large Language Models (LLMs) of the current generation, in contrast, can access enormous amounts of knowledge, encoded (incomprehensibly) in their billions of parameters [4,5,6,7]. Trained correctly, they can recall and combine virtually limitless knowledge from their training set. LLMs have taken the world by storm, and many biomedical researchers already use them in their daily work, for general as well as research tasks [8,9,10]. However, the current way of interacting with LLMs is predominantly manual, virtually non-reproducible, and their behaviour can be erratic. For instance, they are known to confabulate: they make up facts as they go along, and, to make matters worse, are convinced — and convincing — regarding the truth of their confabulations [10,11]. Current efforts towards Artificial General Intelligence have made some progress in addressing these issues by ensembling multiple models [12] with long-term memory stores [13]. However, current AI systems have not yet earned sufficient trust for use in biomedical fields [11]. These areas demand greater care in data privacy, licensing, and transparency than many other issues and cannot be approached without oversight [14].

Computational biomedicine involves many tasks that could be assisted by LLMs, such as experimental design, outcome interpretation, literature evaluation, and web resource exploration. To improve and accelerate these tasks, we have developed BioChatter, a platform optimised for communicating with LLMs in biomedical research (Figure 1). The platform guides the human researcher intuitively through the interaction with the model, while counteracting the problematic behaviours of the LLM. Since the interaction is mainly based on plain text (in any language), it can be used by virtually any researcher.

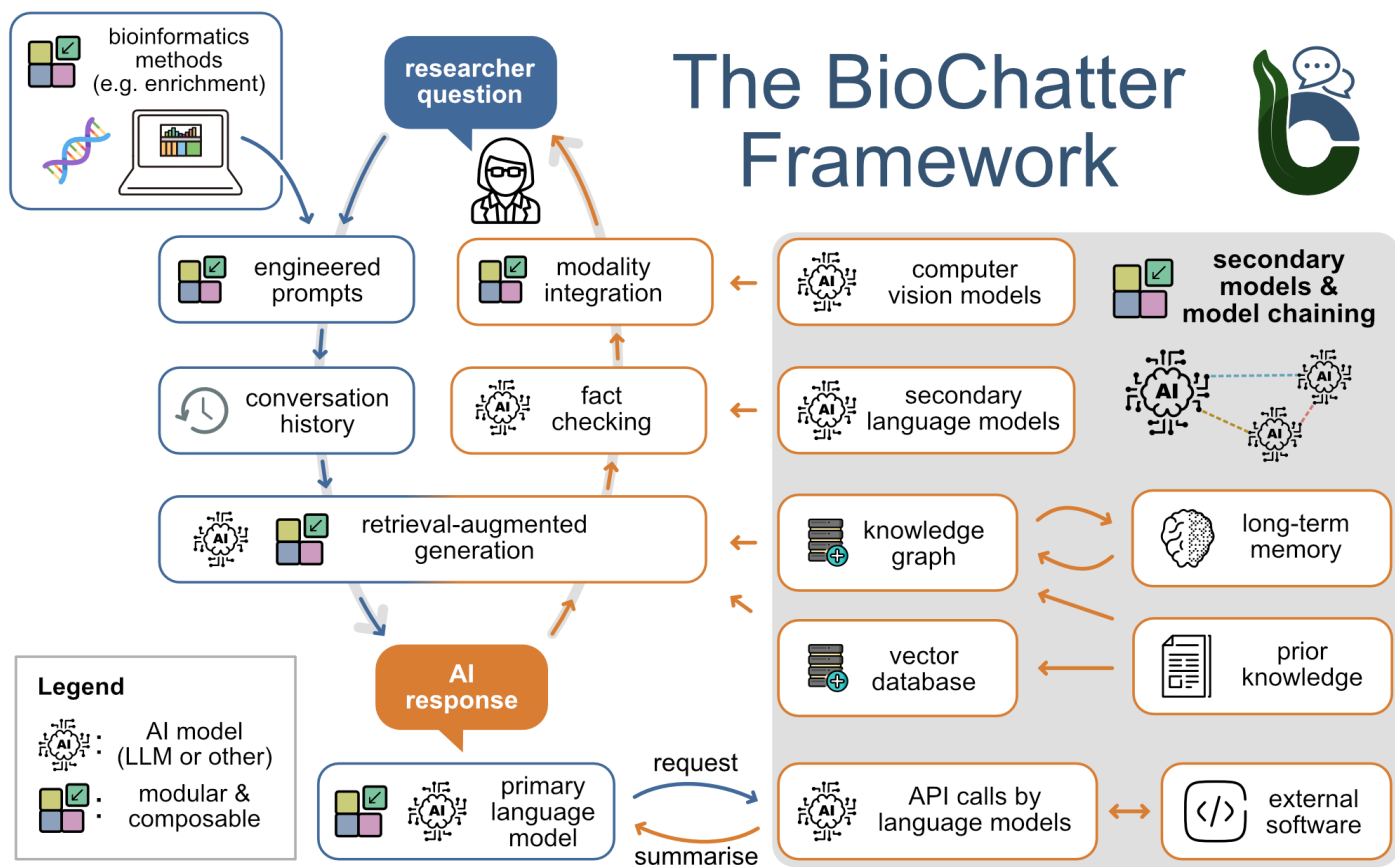

**Figure 1: The BioChatter composable platform architecture (simplified).** LLMs can facilitate many tasks in daily biomedical research practice, for instance interpretation of experimental results or the use of a web resource (top left). BioChatter's main response circuit (blue) composes a number of specifically engineered prompts and passes them (and a conversation history) to the primary LLM, which generates a response for the user based on all inputs. This response is simultaneously used to prompt the secondary circuit (orange), which fulfils auxiliary tasks to complement the primary response. In particular, using search, the secondary circuit queries a database as a prior knowledge repository and compares annotations to the primary response, or uses the knowledge to perform Retrieval-Augmented Generation (RAG). A knowledge graph such as BioCypher [15] can similarly serve as knowledge resource or long-term memory extension of the model. Further, an independent LLM receives the primary response for fact-checking, which can be supplemented with context-specific information by a RAG process. The platform is composable in most aspects, allowing arbitrary extensions to other specialised models for additional tasks orchestrated by the primary LLM.

# Results

BioChatter (https://github.com/biocypher/biochatter) is a Python framework that provides an easy-to-use interface to interact with LLMs and auxiliary technologies via an intuitive API (application programming interface). This way, its functionality can be integrated into any number of user interfaces, such as web apps, command-line interfaces, or Jupyter notebooks (Figure 2).

The framework is designed to be modular: any of its components can be exchanged with other implementations (Figure 1). These functionalities include:

- **basic question-answering** with LLMs hosted by providers (such as OpenAI) as well as locally deployed open-source models

- **reproducible prompt engineering** to guide the LLM towards a specific task or behaviour

- **knowledge graph (KG) querying** with automatic integration of any KG created in the BioCypher framework [15]

- **retrieval-augmented generation** (RAG) using vector database embeddings of user-provided literature

- **model chaining** to orchestrate multiple LLMs and other models in a single conversation using the LangChain framework [16]

- **fact-checking** of LLM responses using a second LLM

- **benchmarking** of LLMs, prompts, and other components

In the following, we briefly describe these components, which are demonstrated in our web apps (https://chat.biocypher.org).

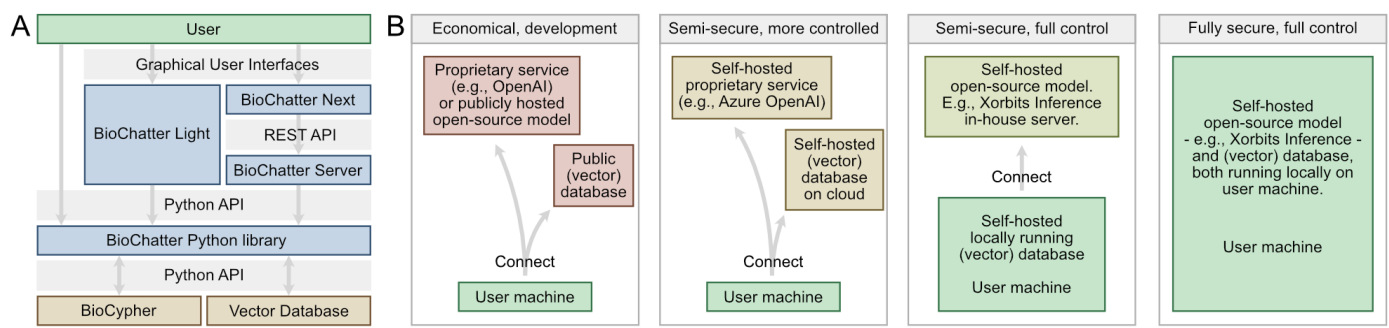

**Figure 2: The BioChatter framework architecture.** A) The BioChatter framework components (blue) connect to knowledge graphs and vector databases (orange). Users (green) can interact with the framework via its Python API, via the lightweight Python frontend using Streamlit (BioChatter Light), or via a fully featured web app with client-server architecture (BioChatter Next). Developers can write simple frontends using the Streamlit framework, or integrate the REST API provided by the BioChatter Server into their own bespoke solutions. B) Different use cases of BioChatter on a spectrum of tradeoff between simplicity/economy (left) and security (right). Economical and simple solutions involve proprietary services that can be used with little effort but are subject to data privacy concerns. Increasingly secure solutions require more effort to set up and maintain, but allow the user to retain more control over their data. Fully local solutions are available given sufficient hardware (starting with contemporary laptops), but are not scalable.

## Question Answering and LLM Connectivity

The core functionality of BioChatter is to interact with LLMs. The framework supports both leading proprietary models such as the GPT series from OpenAI as well as open-source models such as LLaMA2 [17] and Mixtral 8x7B [18] via a flexible open-source deployment framework [19] (see Methods). Currently, the most powerful conversational AI platform, ChatGPT (OpenAI), is surrounded by data privacy concerns [20]. To address this issue, we provide access to the different OpenAI models through their API, which is subject to different, more stringent data protection than the web interface [21], most importantly by disallowing reuse of user inputs for subsequent model training. Further, we aim to preferentially support open-source LLMs to facilitate more transparency in their application and increase data privacy by being able to run a model locally on dedicated hardware and end-user devices [22]. By building on LangChain [16], we support dozens of LLM providers, such as the Xorbits Inference and Hugging Face APIs [19], which can be used to query any of the more than 100 000 open-source models on Hugging Face Hub [23], for instance those on its LLM leaderboard [24]. Although OpenAI's models currently outperform any alternatives in terms of both LLM performance and API convenience, we expect many open-source developments in this area in the future [25]. Therefore, we support plug-and-play exchange of models to enhance biomedical AI readiness, and we implement a bespoke benchmarking framework for the biomedical application of LLMs.

## Prompt Engineering

An essential property of LLMs is their sensitivity to the prompt, i.e., the initial input that guides the model towards a specific task or behaviour. Prompt engineering is an emerging discipline of practical AI, and as such, there are no established best practices [26,27]. Current approaches are mostly trial-and-error-based manual engineering, which is not reproducible and changes with every new model [25]. To address this issue, we include a prompt engineering framework in BioChatter that allows the preservation of prompt sets for specific tasks, which can be shared and reused by the community. In addition, to facilitate the scaling of prompt engineering, we integrate this framework into the benchmarking pipeline, which enables the automated evaluation of prompt sets as new models are published.

## Knowledge Graphs

KGs are a powerful tool to represent and query knowledge in a structured manner. With BioCypher [15], we have developed a framework to create KGs from biomedical data in a user-friendly way while also semantically grounding the data in ontologies. BioChatter is an extension of the BioCypher ecosystem, elevating its user-friendliness further by allowing natural language interactions with the data; any BioCypher KG is automatically compatible with BioChatter. We use information generated in the build process of BioCypher KGs to tune BioChatter's understanding of the data structures and contents, thereby increasing the efficiency of LLM-based KG querying (see Methods). In addition, the ability to connect to any BioCypher KG allows the integration of prior knowledge into the LLM's retrieval, which can be used to ground the model's responses in the context of the KG via in-context learning / retrieval-augmented generation, which can facilitate human-AI interaction via symbolic concepts [7]. We demonstrate the user experience of KG-driven interaction in Supplementary Note 1: Knowledge Graph Retrieval-Augmented Generation and on our website (https://biochatter.org/vignette-kg/).

## Retrieval-Augmented Generation

LLM confabulation is a major issue for biomedical applications, where the consequences of incorrect information can be severe. One popular way of addressing this issue is to apply "in-context learning," which is also more recently referred to as "retrieval-augmented generation" (RAG) [28]. Briefly, RAG relies on injection of information into the model prompt of a pre-trained model and, as such, does not require retraining / fine-tuning; once created, any RAG prompt can be used with any LLM. While this can be done by processing structured knowledge, for instance, from KGs, it is often more efficient to use a semantic search engine to retrieve relevant information from unstructured data sources such as literature. By incorporating the management and integration of vector databases in the BioChatter framework, we allow the user to connect to a vector database, embed an arbitrary number of documents, and then use semantic search to improve the model prompts by adding text fragments relevant to the given question (see Methods). We demonstrate the user experience of RAG in Supplementary Note 2: Retrieval-Augmented Generation and on our website (https://biochatter.org/vignette-rag/).

## Model Chaining and Fact Checking

LLMs cannot only seamlessly interact with human users, but also with other LLMs as well as many other types of models. They understand API calls and can therefore theoretically orchestrate complex multi-step tasks [29,30]. However, implementation is not trivial and the complex process can lead to unpredictable behaviours. We aim to improve the stability of model chaining in biomedical applications by developing bespoke approaches for common biomedical tasks, such as interpretation and design of experiments, evaluating literature, and exploring web resources. While we focus on reusing existing open-source frameworks such as LangChain [16], we also develop bespoke solutions where necessary to provide stability for the given application. As an example, we implemented a fact-checking module that uses a second LLM to evaluate the factual correctness of the primary LLM's responses continuously during the conversation (see Methods).

## Benchmarking

The increasing generality of LLMs poses challenges for their comprehensive evaluation. Specifically, their ability to aid in a multitude of tasks and their great freedom in formatting the answers challenge their evaluation by traditional methods. To circumvent this issue, we focus on specific biomedical tasks and datasets and employ automated validation of the model's responses by a second LLM for advanced assessments. For transparent and reproducible evaluation of LLMs, we implement a benchmarking framework that allows the comparison of models, prompt sets, and all other components of the pipeline. The generic Pytest framework [31] allows for the automated evaluation of a matrix of all possible combinations of components. The results are stored and displayed on our website for simple comparison, and the benchmark is updated upon the release of new models and extensions to the datasets and BioChatter capabilities (https://biochatter.org/benchmark/).

Since the biomedical domain has its own tasks and requirements, we created a bespoke benchmark that allows us to be more precise in the evaluation of components [25]. This is complementary to the existing, general-purpose benchmarks and leaderboards for LLMs [24,32,33]. Furthermore, to prevent leakage of the benchmark data into the training data of the models, a known issue in the general-purpose benchmarks [34], we implemented an encrypted pipeline that contains the benchmark datasets and is only accessible to the workflow that executes the benchmark (see Methods).

Analysis of these benchmarks confirmed the prevailing opinion of OpenAI's leading role in LLM performance (Figure 3 A). Since the benchmark datasets were created to specifically cover functions relevant in BioChatter's application domain, the benchmark results are primarily a measure of the LLMs' usefulness in our applications. OpenAI's GPT models (gpt-4 and gpt-3.5-turbo) lead by some margin on overall performance and consistency, but several open-source models reach high performance in specific tasks. Of note, while the newer version (0125) of gpt-3.5-turbo outperforms the previous version (0613) of gpt-4, version 0125 of gpt-4 shows a significant drop in performance. The performance of open-source models appears to depend on their quantisation level, i.e., the bit-precision used to represent the model's parameters. For models that offer quantisation options, performance apparently plateaus or even decreases after the 4- or 5-bit mark (Figure 3 A). There is no apparent correlation between model size and performance (Pearson's r = 0.171, p < 0.001).

To evaluate the benefit of BioChatter functionality, we compared the performance of models with and without the use of BioChatter's prompt engine for KG querying. The models without prompt engine still have access to the BioCypher schema definition, which details the KG structure, but they do not use the multi-step procedure available through BioChatter. Consequently, the models without prompt engine show a lower performance in creating correct queries than the same models with prompt engine (0.444±0.11 vs. 0.818±0.11, unpaired t-test P < 0.001, Figure 3 B).

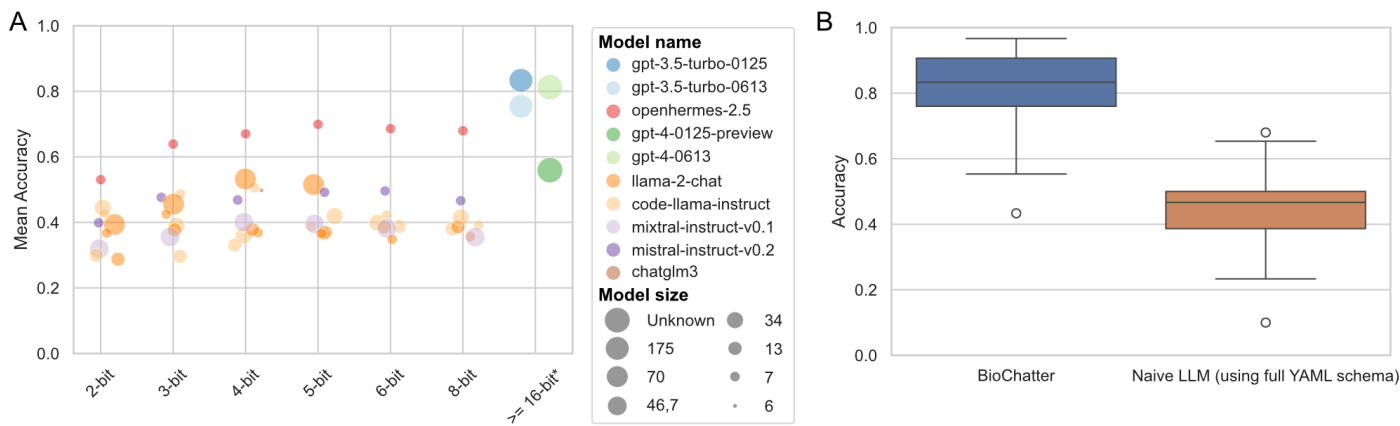

**Figure 3: Benchmark results.** A) Performance of different LLMs (indicated by colour) on the BioChatter benchmark datasets; the y-axis value indicates the average performance across all tasks for each model/size. X-axis jittered for better visibility. While the closed-source models from OpenAI mostly show highest performance, open-source models can perform comparably, but show high variance. Measured performance does not seem to correlate with size (indicated by point size) and quantisation (bit-precision) of the models. *: Of note, many characteristics of OpenAI models are not public, and thus their bit-precision (as well as the exact size of gpt-4) is subject to speculation. B) Comparison of the two benchmark tasks for KG querying show the superior performance of BioChatter's prompt engine (0.818±0.11 vs. 0.444±0.11, unpaired t-test P < 0.001). The test includes all models, sizes, and quantisation levels, and the performance is measured as the average of the two tasks. The BioChatter variant involves a multi-step procedure of constructing the query, while the "naive" version only receives the complete schema definition of the BioCypher KG (which BioChatter also uses as a basis for the prompt engine). The general instructions for both variants are the same, otherwise.

## Discussion

The fast pace of developments around current-generation LLMs poses a great challenge to society as a whole and the biomedical community in particular [35,36,37]. While the potential of these models is enormous, their application is not straightforward, and their use requires a certain level of expertise [38]. In addition, biomedical research is often performed in a siloed way due to the complexity of the domain and systemic incentives that work against open science and collaboration [39,40]. Inspired by the productivity of open source libraries such as LangChain [16], we propose an open framework that allows biomedical researchers to focus on the application of LLMs as opposed to engineering challenges. To keep the framework effective and sustainable, we reuse existing open-source libraries and tools while adapting the advancements from the wider LLM community to the biomedical domain. The transparency we emphasise at every step of the framework is essential to a sustainable application of LLMs in biomedical research and beyond [35].

Efficient human-AI interaction may require a "lingua franca": symbolic representations of concepts at least at the surface level of the conversation [7]. We enable interaction with LLMs on a symbolic level by providing ontological grounding via the synergy of BioChatter with BioCypher KGs. The configuration of BioCypher KGs allows the user to specify the contextual domain by mapping KG concepts to existing ontologies and custom terminology. This way, we guarantee an overlap in the contextual understanding of user and LLM despite the generic nature of most pre-trained models.

We emphasise robustness and objective evaluation of LLM behaviour and performance in interaction with other parts of the framework. We achieve this goal by implementing a living benchmarking framework that allows the automated evaluation of LLMs, prompts, and other components (https://biochatter.org/benchmark/). Even the most recent biomedicine-specific benchmarking efforts are small-scale manual approaches that do not consider the full matrix of possible combinations of components, and many benchmarks are performed by accessing web interfaces of LLMs, which obfuscates important parameters such as model version and temperature [25]. As such, a framework is a necessary step towards the objective and reproducible evaluation of LLMs. We prevent data leakage from the benchmark datasets into the training data of new models by encryption, which is essential for the sustainability of the benchmark as new models are released. The living benchmark will be updated with new questions and tasks as they arise in the community.

The benchmark's results provide convenient selection criteria and a starting point for exploring why some models perform differently than expected. For instance, the benchmark allowed immediate flagging of the drop in performance from the older (0613) to the newer (0125) version of gpt-4. It also identified a range of pre-trained open-source models suitable for our uses, most notably, the openhermes-2.5 model in 4- or 5-bit quantisation. This model is a fine-tuned (on GPT-4-generated data) variant of Mistral 7B v0.1, whose vanilla variants perform considerably worse in our benchmarks. Of note, BioChatter was developed using gpt-3.5-turbo-0613 and, to a lesser extent, gpt-4-0613 and llama-2-chat (13B); the benchmark performance of, for instance, openhermes-2.5 and the newer GPT models thus has not been influenced by BioChatter development.

We facilitate access to LLMs by enabling the use of both proprietary and open-source models, and we provide a flexible deployment framework for the latter. Proprietary models are currently the most economical solution for accessing state-of-the-art models and, as such, they are suitable for users just starting out or lacking the resources to deploy their own models. In contrast, open-source models are quickly catching up in terms of performance [25] and are essential for the sustainability of the field [35]. We allow self-hosting of open-source models on any scale, from dedicated hardware with GPUs, to local deployment on end-user laptops, to browser-based deployment using web technology.

## Limitations

The current generation of LLMs is not yet ready for unsupervised use in biomedical research with its vast array of unique subfields. Effectively supporting this diversity through robust and contextually aware LLM interactions is a daunting task. While we have taken steps to mitigate the risks of using LLMs such as independent benchmarks, fact-checking, and RAG processes, we cannot guarantee that the models will not produce harmful outputs. We see current LLMs, particularly in the scope of the BioCypher ecosystem, as helpful tools to assist human researchers, alleviating menial and repetitive tasks and helping with technical aspects such as query languages. They are not meant to replace human ingenuity and expertise but to augment it with their complementary strengths. Despite the user-friendly design of BioChatter, there may be a learning curve for researchers unfamiliar with LLMs or the specific functionalities of the framework. For maximising its benefit to the community, encouraging adoption and providing adequate training and support will be critical.

## Future directions

Multitask learners that can synthesise, for instance, language, vision, and molecular measurements are an emerging field of research [41,42,43]. Autonomous agents for trivial tasks have already been developed on the basis of LLMs, and we expect this field to mature in the future [30]. As research on multimodal learning and agent behaviour progresses, we plan to integrate these developments into the BioChatter framework. All framework developments will be performed in light of the ethical implications of LLMs, and we will continue to support the use of open-source models to increase transparency and data privacy. While we focus on the biomedical field, the concept of our frameworks can easily be extended to other scientific domains by adjusting domain-specific prompts and data inputs, which are accessible in a composable and user-friendly manner in our frameworks [15]. Our Python library is developed openly on GitHub (https://github.com/biocypher/biochatter) and can be integrated into any downstream user interface solution. We develop under the permissive MIT licence and encourage contributions and suggestions from the community with regard to the addition of bioinformatics tool integrations, prompt engineering, benchmarking, and any other feature.

# (Supplementary / Online) Methods

BioChatter (version 0.4.7 at the time of publication) is a Python library, supporting Python 3.10-3.12, which we ensure with a continuous integration pipeline on GitHub (https://github.com/biocypher/biochatter). We provide documentation at https://biochatter.org, including a tutorial and API reference. All packages are developed openly and according to modern standards of software development [44]; we use the permissive MIT licence to encourage downstream use and development. We include a code of conduct and contributor guidelines to offer accessibility and inclusivity to all who are interested in contributing to the framework.

## Applications

To demonstrate basic and advanced use cases of the framework, we provide two web apps, BioChatter Light and BioChatter Next.

BioChatter Light is a web app based on the Streamlit framework (version 1.31.1, https://streamlit.io), which is written in Python and can be deployed locally or on a server (https://github.com/biocypher/biochatter-light). The ease with which Streamlit allows the creation of interactive web apps in pure Python enables rapid iteration and agile development of new features, with the tradeoff of limited customisation and scalability. This framework is suitable for rapid prototyping of bespoke solutions for specific use cases. For an up-to-date overview and preview of the current functionality of the platform, please visit the online preview.

BioChatter Next (https://github.com/biocypher/biochatter-next) is a modern web app with server-client architecture, based on the open-source template of ChatGPT-Next-Web (https://github.com/ChatGPTNextWeb/ChatGPT-Next-Web). It is written combining Typescript and Python and uses Next.js (v13.4.9) for a sleek frontend and Flask (v3.0.0) as backend. It demonstrates the use of BioChatter in a modern web app, including full customisation and scalability and localisation in 18 languages. However, this comes at the cost of increased complexity and development time. To provide seamless integration of the BioChatter backend into existing frontend solutions, we provide the server implementation at https://github.com/biocypher/biochatter-server and as a Docker image in our Docker Hub organisation (https://hub.docker.com/repository/docker/biocypher/biochatter-server).

We invite all interested researchers to select the framework that best suits their needs, or use the BioChatter server or library in their existing solutions.

## Benchmarking

The benchmarking framework examines a matrix of component combinations using the parameterisation feature of Pytest [31]. This implementation allows for the automated evaluation of all possible combinations of components, such as LLMs, prompts, and datasets. We performed the benchmarks on a MacBook Pro with an M3 Max chip with 40-core GPU and 128GB of RAM. As a default, we ran each test five times to account for the stochastic nature of LLMs. We generally set the temperature to the lowest value possible for each model to decrease fluctuation.

The Pytest matrix uses a hash-based system to evaluate whether a model-dataset combination has been run before. Briefly, the hash is calculated from the dictionary representation of the test parameters, and the test is skipped if the combination of hash and model name is already present in the database. This hashing optimises for efficiency by only running modified or newly added tests. The individual dimensions of the matrix are:

- **LLMs**: Testing proprietary (OpenAI) and open-source models (commonly using the Xorbits Inference API and HuggingFace models) against the same set of tasks is the primary aim of our benchmarking framework. We facilitate the automation of testing by including a programmatic way of deploying open-source models.

- **prompts**: Since model performance can dramatically rely on the used prompts, a set of prompts for each task with varying degrees of specificity and fixed as well as variable components is used to evaluate this variability.

- **datasets**: We test various tasks using a set of datasets for each task in question-answer-style.

- **data processing**: Some data processing steps can have great impact on the downstream performance of LLMs. For instance, we test the conversion of numbers (which LLMs are notoriously bad at handling) to categorical text (e.g., low, medium, high).

- **model quantisations**: We test a set of quantisations for each model (where available) to account for the trade-off between model size, inference speed, and performance.

- **model parameters**: Where suitable, we test a set of parameters for each model, such as "temperature," which determines the reproducibility of model responses.

- **integrations**: We write dedicated tests for specific tasks that require integrations, for instance with knowledge graphs or vector databases.

- **stochasticity**: To account for variability in model responses, we include a parameter to run each test multiple times and generate summary statistics.

- **sentiment and behaviour**: To assess whether the models exhibit the desired behaviour patterns for each of the personas, we let a second LLM evaluate the responses based on a set of criteria, including professionalism and politeness.

The Pytest framework is implemented at https://github.com/biocypher/biochatter/blob/main/benchmark, and more information is available at https://biochatter.org/benchmarking. The benchmark is updated upon the release of new models and extensions to the datasets, and continuously available at https://biochatter.org/benchmark. We will run the benchmark on new models and variants (including fine-tuned models) upon requests from the community, which can be made on GitHub using our issue template (https://github.com/biocypher/biochatter/issues/new/choose). The living benchmark process is inspired by test-driven development, meaning test cases are created based on specific features or behaviors that are desired. When a model doesn't initially produce the optimal response, which is often the case, adjustments are made to various elements of the framework, including prompts or functions, to enhance the model's effectiveness. Monitoring the model's performance on these tests over time allows us to assess the framework's reliability and pinpoint areas that need improvement.

To prevent leakage of benchmarking data (and subsequent contamination of future LLMs), we implement an encryption routine on the benchmark datasets. The encryption is performed using a hybrid encryption scheme, where the data are encrypted with a symmetric key, which is in turn encrypted with an asymmetric key. The datasets are stored in a dedicated encrypted pipeline that is only accessible to the workflow that executes the benchmark. These processes are implemented at https://github.com/biocypher/llm-test-dataset and accessed from the benchmark procedure in BioChatter.

## Knowledge Graphs

We utilise the close connection between BioChatter and the BioCypher framework [15] to integrate knowledge graph (KG) queries into the BioChatter API. In the BioCypher KG creation, we use a configuration file to map KG contents to ontology terms, including information about each of the entities. For instance, we detail the properties of a node and the source and target classes of an edge. Additionally, during the KG build process, we enrich this information and save it to a YAML file and, optionally, directly to the KG. This information is used by BioChatter to tune its understanding of the KG, which allows the LLM to query the KG more efficiently.

By understanding the context of the KG, the exact contents, and the exact spelling of all identifiers and properties, we effectively support the LLM in generating correct queries. The query generation process is broken up into multiple steps by BioChatter: recognising entities and relationships according to the user's question, estimating properties to be used in the query, and generating a syntactically correct query in the query language of the database, based on the results from the previous steps and constraints given by the KG schema information. This procedure is implemented in the `prompts.py` module. To evaluate the quality of this process, we dedicate a module in the benchmark to the query generation process with a range of questions and KG schemata.

To illustrate the usage of this feature, we provide a demonstration repository at https://github.com/biocypher/pole including a KG build procedure and an instance of BioChatter Light, which can be run using a single Docker Compose command. The pole KG can also be used in conjunction with the BioChatter Next app by using the `docker-compose.yaml` file to build the application locally. A demonstration of this use case is available in Supplementary Note 1: Knowledge Graph Retrieval-Augmented Generation and on our website (https://biochatter.org/vignette-kg/).

## Retrieval-Augmented Generation

While current LLMs possess extensive internal general knowledge, they may not know how to prioritise very specific scientific results, or they may not have had access to some research articles in their training data (e.g., due to their recency or licensing issues). To bridge this gap, we can provide additional information from relevant publications to the model via the prompt. However, we frequently cannot add entire publications to the prompt, since the input length of current models is still restricted; we need to isolate the information that is specifically relevant to the question given by the user. To find this information, we perform a semantic similarity search between the user's question and the contents of user-provided scientific articles (or other texts). The most efficient way to do this mapping is by using a vector database [45].

The contextual background information provided by the user (e.g., by uploading a scientific article of prior work related to the experiment to be interpreted) is split into pieces suitable to be digested by the LLM, which are individually embedded by the model. These embeddings (represented by vectors) are used to store the text fragments in a vector database; the storage as vectors allows fast and efficient retrieval of similar entities via the comparison of individual vectors. For example, the two sentences "Amyloid beta levels are associated with Alzheimer's Disease stage." and "One of the most important clinical markers of AD progression is the amount of deposited A-beta 42." would be closely associated in a vector database (given the embedding model is of sufficient quality, i.e., similar to GPT-3 or better), while traditional text-based similarity metrics probably would not identify them as highly similar.

By comparing the user's question to prior knowledge in the vector database, we can extract the relevant pieces of information from the entire background. Even better, we can first use an LLM to generate an answer to the user's question and then use this answer to query the vector database for relevant information. Regardless of whether the initial answer is correct, it is likely that the "fake answer" is more semantically similar to the relevant pieces of information than the user's question [45]. Semantic search results (for instance, single sentences directly related to the topic of the question) are then sufficiently small to be added to the prompt. In this way, the model can learn from additional context without the need for retraining or fine-tuning. This method is sometimes described as in-context learning [28] or retrieval-augmented generation [46].

To provide access to this functionality in BioChatter, we implement classes for the connection to, and management of, vector database systems (in the `vectorstore.py` module), and for performing semantic search on the vector database and injecting the results into the prompt (in the `vectorstore_agent.py` module). An analogous implementation for KG retrieval is available in the `database_agent.py` module. Both retrieval mechanisms are integrated and provided to the BioChatter API via the `rag_agent.py` module. To demonstrate the use of the API, we add a "Retrieval-Augmented Generation" tab to the preview apps that allows the upload of text documents to be added to a vector database, which then can be queried to add contextual information to the prompt sent to the primary model. This contextual information is transparently displayed. Since this functionality requires a connection to a vector database system, we provide connectivity to a Milvus service, including a way to start the service in conjunction with a BioCypher knowledge graph and the BioChatter Light app in one Docker Compose workflow.

An example use case of this functionality is available in Supplementary Note 2: Retrieval-Augmented Generation and on our website (https://biochatter.org/vignette-rag/).

## Deployment of Open-Source Models

To facilitate access to open-source models, we adopt a flexible deployment framework based on the Xorbits Inference API [19]. Xorbits Inference includes a large number of open-source models out of the box, and new models from Hugging Face Hub [47] can be added using the intuitive graphical user interface. We used Xorbits Inference version 0.8.4 to deploy the benchmarked models, and we provide a Docker Compose repository to deploy the app on a Linux server with Nvidia GPUs (https://github.com/biocypher/xinference-docker-builtin/). This Compose uses the multi-architecture image (for ARM64 and AMD64 chips) we provide on our Docker Hub organisation (https://hub.docker.com/repository/docker/biocypher/xinference-builtin). On Mac OS with Apple Silicon chips, Docker does not have access to the GPU driver, and as such, Xinference needs to be deployed natively.

## Model Chaining

The ability of LLMs to control external software, including other LLMs, opens up a wide range of possibilities for the orchestration of complex tasks. A simple example is the implementation of a correcting agent, which receives the output of the primary model and checks it for factual correctness. If the agent detects an error, it can prompt the primary model to correct its output, or forward this correction to the user directly. Since this relies on the internal knowledge base of the correcting agent, the same caveats apply, as the correcting agent may confabulate as well. However, since the agent is independent of the primary model (being set up with dedicated prompts), it is less likely to confabulate in the same way.

This approach can be extended to a more complex model chain, where the correcting agent, for example, can query a knowledge graph or a vector database to ground its responses in prior knowledge. These chains are easy to implement, and some are available out of the box in the LangChain framework [16]. However, they can behave unpredictably, which increases with the number of links in the chain and, as such, should be tightly controlled. They also add to the computational burden of the system, which is particularly relevant for deployments on end-user devices.

## Author Contributions

SL conceptualised and developed the platform and wrote the manuscript. SF integrated BioChatter continuous integration pipelines and developed both front-end and back-end components for the BioChatter Next server. The BioChatter consortium members contributed to the development of the platform and provided feedback on the manuscript. AM implemented the local deployment functionality. CW architected the BioChatter Next server infrastructure. JB provided guidance and supervision as well as hardware resources for local LLM use and contributed to performance benchmarking. NK implemented benchmarking procedures. QM oversaw the development and deployment of the BioChatter Next server environment. JSR supervised the project, revised the manuscript, and acquired funding. All authors read and approved the final manuscript.

## Acknowledgements

We thank Hanna Schumacher, Daniel Dimitrov, Pau Badia i Mompel, and Aurelien Dugourd for feedback on the original draft of the manuscript and the software.

This work was supported by funding from the European Union's Horizon 2020 research and innovation programme under grant agreement No 965193 (JSR), award U54-AG075931 (QM) from the National Institutes of Health, award NSF1945971 (QM) from the National Science Foundation, and the Pelotonia Institute of Immuno-Oncology.

## Disclaimer

This project is funded by the European Union under grant agreement No. 101057619. Views and opinions expressed are however those of the author(s) only and do not necessarily reflect those of the European Union or European Health and Digital Executive Agency (HADEA). Neither the European Union nor the granting authority can be held responsible for them. This work was also partly supported by the Swiss State Secretariat for Education, Research and Innovation (SERI) under contract No. 22.00115.

## Conflict of Interest

JSR reports funding from GSK, Pfizer and Sanofi and fees/honoraria from Travere Therapeutics, Stadapharm, Pfizer, Grunenthal, Owkin, and Astex Pharmaceuticals.

# References

1. **Study reveals cancer's 'infinite' ability to evolve**
   James Gallagher
   *BBC News* (2023-04-12) https://www.bbc.com/news/health-65252510

2. **Current progress and open challenges for applying deep learning across the biosciences**
   Nicolae Sapoval, Amirali Aghazadeh, Michael G Nute, Dinler A Antunes, Advait Balaji, Richard Baraniuk, CJ Barberan, Ruth Dannenfelser, Chen Dun, Mohammadamin Edrisi, … Todd J Treangen
   *Nature Communications* (2022-04-01) https://doi.org/gp26xk
   DOI: 10.1038/s41467-022-29268-7 · PMID: 35365602 · PMCID: PMC8976012

3. **Capacity limits of information processing in the brain**
   René Marois, Jason Ivanoff
   *Trends in Cognitive Sciences* (2005-06) https://doi.org/d5gmqt
   DOI: 10.1016/j.tics.2005.04.010 · PMID: 15925809

4. **PaLM: Scaling Language Modeling with Pathways**
   Aakanksha Chowdhery, Sharan Narang, Jacob Devlin, Maarten Bosma, Gaurav Mishra, Adam Roberts, Paul Barham, Hyung Won Chung, Charles Sutton, Sebastian Gehrmann, … Noah Fiedel
   *arXiv* (2022) https://doi.org/kfxf
   DOI: 10.48550/arxiv.2204.02311

5. **LaMDA: Language Models for Dialog Applications**
   Romal Thoppilan, Daniel De Freitas, Jamie Hall, Noam Shazeer, Apoorv Kulshreshtha, Heng-Tze Cheng, Alicia Jin, Taylor Bos, Leslie Baker, Yu Du, … Quoc Le
   *arXiv* (2022) https://doi.org/kmfc
   DOI: 10.48550/arxiv.2201.08239

6. **GPT-4 Technical Report**
   OpenAI, :, Josh Achiam, Steven Adler, Sandhini Agarwal, Lama Ahmad, Ilge Akkaya, Florencia Leoni Aleman, Diogo Almeida, Janko Altenschmidt, … Barret Zoph
   *arXiv* (2023) https://doi.org/grx4cb
   DOI: 10.48550/arxiv.2303.08774

7. **Symbols as a Lingua Franca for Bridging Human-AI Chasm for Explainable and Advisable AI Systems**
   Subbarao Kambhampati, Sarath Sreedharan, Mudit Verma, Yantian Zha, Lin Guan
   *Proceedings of the AAAI Conference on Artificial Intelligence* (2022-06-28) https://doi.org/gthj8r
   DOI: 10.1609/aaai.v36i11.21488

8. **Autonomous chemical research with large language models**
   Daniil A Boiko, Robert MacKnight, Ben Kline, Gabe Gomes
   *Nature* (2023-12-20) https://doi.org/gs9v4v
   DOI: 10.1038/s41586-023-06792-0 · PMID: 38123806 · PMCID: PMC10733136

9. **Assessing GPT-4 for cell type annotation in single-cell RNA-seq analysis**
   Wenpin Hou, Zhicheng Ji
   *Cold Spring Harbor Laboratory* (2023-04-21) https://doi.org/gsznzg
   DOI: 10.1101/2023.04.16.537094 · PMID: 37131626 · PMCID: PMC10153208

10. **How will generative AI disrupt data science in drug discovery?**
    Jean-Philippe Vert

*Nature Biotechnology* (2023-05-08) <https://doi.org/gsznzd>
DOI: [10.1038/s41587-023-01789-6](https://doi.org/10.1038/s41587-023-01789-6) · PMID: [37156917](https://pubmed.ncbi.nlm.nih.gov/37156917)

11. **Foundation models for generalist medical artificial intelligence**
    Michael Moor, Oishi Banerjee, Zahra Shakeri Hossein Abad, Harlan M Krumholz, Jure Leskovec, Eric J Topol, Pranav Rajpurkar
    *Nature* (2023-04-12) <https://doi.org/gr4td4>
    DOI: [10.1038/s41586-023-05881-4](https://doi.org/10.1038/s41586-023-05881-4) · PMID: [37045921](https://pubmed.ncbi.nlm.nih.gov/37045921)

12. 🦜🔗 **Langchain** <https://python.langchain.com/>

13. **AutoGPT Official**
    AutoGPT Official
    (2024-02-16) <https://autogpt.net/>

14. **Towards Conversational Diagnostic AI**
    Tao Tu, Anil Palepu, Mike Schaekermann, Khaled Saab, Jan Freyberg, Ryutaro Tanno, Amy Wang, Brenna Li, Mohamed Amin, Nenad Tomasev, … Vivek Natarajan
    *arXiv* (2024) <https://doi.org/gtdmpj>
    DOI: [10.48550/arxiv.2401.05654](https://doi.org/10.48550/arxiv.2401.05654)

15. **Democratizing knowledge representation with BioCypher**
    Sebastian Lobentanzer, Patrick Aloy, Jan Baumbach, Balazs Bohar, Vincent J Carey, Pornpimol Charoentong, Katharina Danhauser, Tunca Doğan, Johann Dreo, Ian Dunham, … Julio Saez-Rodriguez
    *Nature Biotechnology* (2023-06-19) <https://doi.org/gszqjr>
    DOI: [10.1038/s41587-023-01848-y](https://doi.org/10.1038/s41587-023-01848-y) · PMID: [37337100](https://pubmed.ncbi.nlm.nih.gov/37337100)

16. **LangChain**
    Harrison Chase
    (2022-10-17) <https://github.com/langchain-ai/langchain>

17. **Llama 2: Open Foundation and Fine-Tuned Chat Models**
    Hugo Touvron, Louis Martin, Kevin Stone, Peter Albert, Amjad Almahairi, Yasmine Babaei, Nikolay Bashlykov, Soumya Batra, Prajjwal Bhargava, Shruti Bhosale, … Thomas Scialom
    *arXiv* (2023) <https://doi.org/ktkj>
    DOI: [10.48550/arxiv.2307.09288](https://doi.org/10.48550/arxiv.2307.09288)

18. **Mixtral of Experts**
    Albert Q Jiang, Alexandre Sablayrolles, Antoine Roux, Arthur Mensch, Blanche Savary, Chris Bamford, Devendra Singh Chaplot, Diego de las Casas, Emma Bou Hanna, Florian Bressand, … William El Sayed
    *arXiv* (2024) <https://doi.org/gtc2g3>
    DOI: [10.48550/arxiv.2401.04088](https://doi.org/10.48550/arxiv.2401.04088)

19. **xorbitsai/inference**
    Xorbits
    (2024-02-17) <https://github.com/xorbitsai/inference>

20. <https://www.reuters.com/technology/european-data-protection-board-discussing-ai-policy-thursday-meeting-2023-04-13/>

21. **Terms of use** <https://openai.com/policies/terms-of-use>

22. **Why open-source generative AI models are an ethical way forward for science**
    Arthur Spirling

*Nature* (2023-04-18) https://doi.org/gsqx6v
DOI: 10.1038/d41586-023-01295-4 · PMID: 37072520

23. **Hugging Face Hub documentation** https://huggingface.co/docs/hub/index

24. **Open LLM Leaderboard - a Hugging Face Space by HuggingFaceH4**
    https://huggingface.co/spaces/HuggingFaceH4/open_llm_leaderboard

25. **BioLLMBench: A Comprehensive Benchmarking of Large Language Models in Bioinformatics**
    Varuni Sarwal, Viorel Munteanu, Timur Suhodolschi, Dumitru Ciorba, Eleazar Eskin, Wei Wang, Serghei Mangul
    *Cold Spring Harbor Laboratory* (2023-12-20) https://doi.org/gtbgvk
    DOI: 10.1101/2023.12.19.572483

26. **A Prompt Pattern Catalog to Enhance Prompt Engineering with ChatGPT**
    Jules White, Quchen Fu, Sam Hays, Michael Sandborn, Carlos Olea, Henry Gilbert, Ashraf Elnashar, Jesse Spencer-Smith, Douglas C Schmidt
    *arXiv* (2023) https://doi.org/grxct8
    DOI: 10.48550/arxiv.2302.11382

27. **Principled Instructions Are All You Need for Questioning LLaMA-1/2, GPT-3.5/4**
    Sondos Mahmoud Bsharat, Aidar Myrzakhan, Zhiqiang Shen
    *arXiv* (2023) https://doi.org/gtdnfg
    DOI: 10.48550/arxiv.2312.16171

28. **HuggingGPT: Solving AI Tasks with ChatGPT and its Friends in Hugging Face**
    Yongliang Shen, Kaitao Song, Xu Tan, Dongsheng Li, Weiming Lu, Yueting Zhuang
    *arXiv* (2023) https://doi.org/gskd97
    DOI: 10.48550/arxiv.2303.17580

29. **Gorilla: Large Language Model Connected with Massive APIs**
    Shishir G Patil, Tianjun Zhang, Xin Wang, Joseph E Gonzalez
    *arXiv* (2023) https://doi.org/gtbgvm
    DOI: 10.48550/arxiv.2305.15334

30. **A Survey on Large Language Model based Autonomous Agents**
    Lei Wang, Chen Ma, Xueyang Feng, Zeyu Zhang, Hao Yang, Jingsen Zhang, Zhiyuan Chen, Jiakai Tang, Xu Chen, Yankai Lin, … Ji-Rong Wen
    *arXiv* (2023) https://doi.org/gsv93m
    DOI: 10.48550/arxiv.2308.11432

31. **pytest-dev/pytest**
    pytest-dev
    (2024-02-17) https://github.com/pytest-dev/pytest

32. **Large language models encode clinical knowledge**
    Karan Singhal, Shekoofeh Azizi, Tao Tu, SSara Mahdavi, Jason Wei, Hyung Won Chung, Nathan Scales, Ajay Tanwani, Heather Cole-Lewis, Stephen Pfohl, … Vivek Natarajan
    *Nature* (2023-07-12) https://doi.org/gsgp8c
    DOI: 10.1038/s41586-023-06291-2 · PMID: 37438534 · PMCID: PMC10396962

33. **Holistic Evaluation of Language Models (HELM)** https://crfm.stanford.edu/helm/lite/latest/

34. **NLP Evaluation in trouble: On the Need to Measure LLM Data Contamination for each Benchmark**

Oscar Sainz, Jon Ander Campos, Iker García-Ferrero, Julen Etxaniz, Oier Lopez de Lacalle, Eneko Agirre
*arXiv* (2023) <https://doi.org/gtbgvp>
DOI: [10.48550/arxiv.2310.18018](https://doi.org/10.48550/arxiv.2310.18018)

35. **There are holes in Europe's AI Act — and researchers can help to fill them**
Nature
(2024-01-10) <https://doi.org/gtdnfb>
DOI: [10.1038/d41586-024-00029-4](https://doi.org/10.1038/d41586-024-00029-4) · PMID: [38200306](https://pubmed.ncbi.nlm.nih.gov/38200306)

36. **Is AI leading to a reproducibility crisis in science?**
Philip Ball
*Nature* (2023-12-05) <https://doi.org/gs8bmk>
DOI: [10.1038/d41586-023-03817-6](https://doi.org/10.1038/d41586-023-03817-6) · PMID: [38052897](https://pubmed.ncbi.nlm.nih.gov/38052897)

37. **Generative AI could revolutionize health care — but not if control is ceded to big tech**
Augustin Toma, Senthujan Senkaiahliyan, Patrick R Lawler, Barry Rubin, Bo Wang
*Nature* (2023-11-30) <https://doi.org/gtdnd9>
DOI: [10.1038/d41586-023-03803-y](https://doi.org/10.1038/d41586-023-03803-y) · PMID: [38036861](https://pubmed.ncbi.nlm.nih.gov/38036861)

38. **The perpetual motion machine of AI-generated data and the distraction of ChatGPT as a 'scientist'**
Jennifer Listgarten
*Nature Biotechnology* (2024-01-25) <https://doi.org/gtf6xh>
DOI: [10.1038/s41587-023-02103-0](https://doi.org/10.1038/s41587-023-02103-0) · PMID: [38273064](https://pubmed.ncbi.nlm.nih.gov/38273064)

39. **Scientific Utopia**
Brian A Nosek, Jeffrey R Spies, Matt Motyl
*Perspectives on Psychological Science* (2012-11) <https://doi.org/f4fc2k>
DOI: [10.1177/1745691612459058](https://doi.org/10.1177/1745691612459058) · PMID: [26168121](https://pubmed.ncbi.nlm.nih.gov/26168121) · PMCID: [PMC10540222](https://www.ncbi.nlm.nih.gov/pmc/articles/PMC10540222)

40. **Open science — embrace it before it's too late**
Nature
(2024-02-06) <https://doi.org/gthbjn>
DOI: [10.1038/d41586-024-00322-2](https://doi.org/10.1038/d41586-024-00322-2) · PMID: [38321141](https://pubmed.ncbi.nlm.nih.gov/38321141)

41. **Git-Theta: A Git Extension for Collaborative Development of Machine Learning Models**
Nikhil Kandpal, Brian Lester, Mohammed Muqeeth, Anisha Mascarenhas, Monty Evans, Vishal Baskaran, Tenghao Huang, Haokun Liu, Colin Raffel
*arXiv* (2023) <https://doi.org/gtdnfd>
DOI: [10.48550/arxiv.2306.04529](https://doi.org/10.48550/arxiv.2306.04529)

42. **Crosslingual Generalization through Multitask Finetuning**
Niklas Muennighoff, Thomas Wang, Lintang Sutawika, Adam Roberts, Stella Biderman, Teven Le Scao, MSaiful Bari, Sheng Shen, Zheng-Xin Yong, Hailey Schoelkopf, … Colin Raffel
*arXiv* (2022) <https://doi.org/gtdnfc>
DOI: [10.48550/arxiv.2211.01786](https://doi.org/10.48550/arxiv.2211.01786)

43. **MiniGPT-v2: large language model as a unified interface for vision-language multi-task learning**
Jun Chen, Deyao Zhu, Xiaoqian Shen, Xiang Li, Zechun Liu, Pengchuan Zhang, Raghuraman Krishnamoorthi, Vikas Chandra, Yunyang Xiong, Mohamed Elhoseiny
*arXiv* (2023) <https://doi.org/gtdnff>
DOI: [10.48550/arxiv.2310.09478](https://doi.org/10.48550/arxiv.2310.09478)

44. **The TRUST Principles for digital repositories**

Dawei Lin, Jonathan Crabtree, Ingrid Dillo, Robert R Downs, Rorie Edmunds, David Giaretta, Marisa De Giusti, Hervé L'Hours, Wim Hugo, Reyna Jenkyns, … John Westbrook
*Scientific Data* (2020-05-14) https://doi.org/ggwrtj
DOI: 10.1038/s41597-020-0486-7 · PMID: 32409645 · PMCID: PMC7224370

45. **Large Language Models for Information Retrieval: A Survey**
    Yutao Zhu, Huaying Yuan, Shuting Wang, Jiongnan Liu, Wenhan Liu, Chenlong Deng, Haonan Chen, Zhicheng Dou, Ji-Rong Wen
    *arXiv* (2023) https://doi.org/gtbgvn
    DOI: 10.48550/arxiv.2308.07107

46. **Retrieval-Augmented Generation for Knowledge-Intensive NLP Tasks**
    Patrick Lewis, Ethan Perez, Aleksandra Piktus, Fabio Petroni, Vladimir Karpukhin, Naman Goyal, Heinrich Küttler, Mike Lewis, Wen-tau Yih, Tim Rocktäschel, … Douwe Kiela
    *Advances in Neural Information Processing Systems* (2020)
    https://proceedings.neurips.cc/paper_files/paper/2020/file/6b493230205f780e1bc26945df7481e5-Paper.pdf

47. **Hugging Face – The AI community building the future.** (2024-02-13) https://huggingface.co/

48. **ABC transporters affects tumor immune microenvironment to regulate cancer immunotherapy and multidrug resistance**
    Jingyi Fan, Kenneth Kin Wah To, Zhe-Sheng Chen, Liwu Fu
    *Drug Resistance Updates* (2023-01) https://doi.org/gtg7tg
    DOI: 10.1016/j.drup.2022.100905 · PMID: 36463807

49. **ABC transporters and the hallmarks of cancer: roles in cancer aggressiveness beyond multidrug resistance**
    Wanjiru Muriithi, Lucy Wanjiku Macharia, Carlos Pilotto Heming, Juliana Lima Echevarria, Atunga Nyachieo, Paulo Niemeyer Filho, Vivaldo Moura Neto
    *Cancer Biology and Medicine* (2020) https://doi.org/gtg7tj
    DOI: 10.20892/j.issn.2095-3941.2019.0284 · PMID: 32587767 · PMCID: PMC7309456

50. **Advances in the structure, mechanism and targeting of chemoresistance-linked ABC transporters**
    Andaleeb Sajid, Hadiar Rahman, Suresh V Ambudkar
    *Nature Reviews Cancer* (2023-09-15) https://doi.org/gss2df
    DOI: 10.1038/s41568-023-00612-3 · PMID: 37714963

51. **ATP-binding cassette (ABC) transporters in cancer: A review of recent updates**
    Jing-Quan Wang, Zhuo-Xun Wu, Yuqi Yang, Qiu-Xu Teng, Yi-Dong Li, Zi-Ning Lei, Khushboo A Jani, Neeraj Kaushal, Zhe-Sheng Chen
    *Journal of Evidence-Based Medicine* (2021-08-13) https://doi.org/gtg7th
    DOI: 10.1111/jebm.12434 · PMID: 34388310

## The BioChatter Consortium

The BioChatter Consortium is a community of researchers and developers who are interested in the application of LLMs to biomedical research. We list here alphabetically all members of the consortium who have contributed to the development of the BioChatter framework.

- Adrián G. Díaz (Interuniversity Institute of Bioinformatics in Brussels, Brussels, Belgium and Structural Biology Brussels, Vrije Universiteit Brussels, Brussels, Belgium)

- Andreas Maier (Institute for Computational Systems Biology, University of Hamburg, Hamburg, Germany)
- Aurelien Dugourd (Heidelberg University, Faculty of Medicine and Heidelberg University Hospital, Institute for Computational Biomedicine, Heidelberg, Germany)
- Cankun Wang (Department of Biomedical Informatics, The Ohio State University, Columbus, Ohio, USA)
- Emma Verkinderen (Interuniversity Institute of Bioinformatics in Brussels, Université Libre de Bruxelles-Vrije Universiteit Brussel, 1050, Brussels, Belgium)
- Fernando M. Delgado-Chaves (Institute for Computational Systems Biology, University of Hamburg, Hamburg, Germany)
- Jan Baumbach (Institute for Computational Systems Biology, University of Hamburg, Hamburg, Germany and Computational Biomedicine Lab, Department of Mathematics and Computer Science, University of Southern Denmark, Odense, Denmark)
- Julio Saez-Rodriguez (Heidelberg University, Faculty of Medicine and Heidelberg University Hospital, Institute for Computational Biomedicine, Heidelberg, Germany)
- Megan McNutt (Department of Biomedical Informatics, The Ohio State University, Columbus, Ohio, USA)
- Nils Krehl (Heidelberg University, Faculty of Medicine and Heidelberg University Hospital, Institute for Computational Biomedicine, Heidelberg, Germany)
- Qin Ma (Department of Biomedical Informatics, The Ohio State University, Columbus, Ohio, USA)
- Sebastian Beier (Institute of Bio- and Geosciences (IBG-4: Bioinformatics), Bioeconomy Science Center (BioSC), CEPLAS, Forschungszentrum Jülich, Jülich, Germany)
- Sebastian Lobentanzer (Heidelberg University, Faculty of Medicine and Heidelberg University Hospital, Institute for Computational Biomedicine, Heidelberg, Germany)
- Shaohong Feng (Department of Biomedical Informatics, The Ohio State University, Columbus, Ohio, USA)
- Trang T. Le (Bristol Myers Squibb, 250 Water St, Cambridge, MA 02141, USA)
- Xiao-Ran Zhou (Institute of Bio- and Geosciences (IBG-4: Bioinformatics), Bioeconomy Science Center (BioSC), CEPLAS, Forschungszentrum Jülich, Jülich, Germany)
- Yuyao Song (European Molecular Biology Laboratory-European Bioinformatics Institute (EMBL-EBI), Wellcome Genome Campus, Hinxton, United Kingdom)

# Supplementary Materials

## Supplementary Note 1: Knowledge Graph Retrieval-Augmented Generation

This vignette demonstrates the KG module of BioChatter as used by the BioChatter Next application. It is available online (including video and more information) at https://biochatter.org/vignette-kg/. We connect to a BioCypher knowledge graph (KG) to retrieve relevant information for a given question. We then use the retrieved information to generate a response to the question. The application can connect to any real-world BioCypher KG by providing the connection details in the `KG Settings` dialog.

**Background**

For the demonstration purposes of this vignette, we include a demo KG based on an open-source dataset of crime statistics in Manchester, because it allows us to redistribute the KG due to its small size and public domain licence, and because it is easily understood. This is the schema of the demo KG:

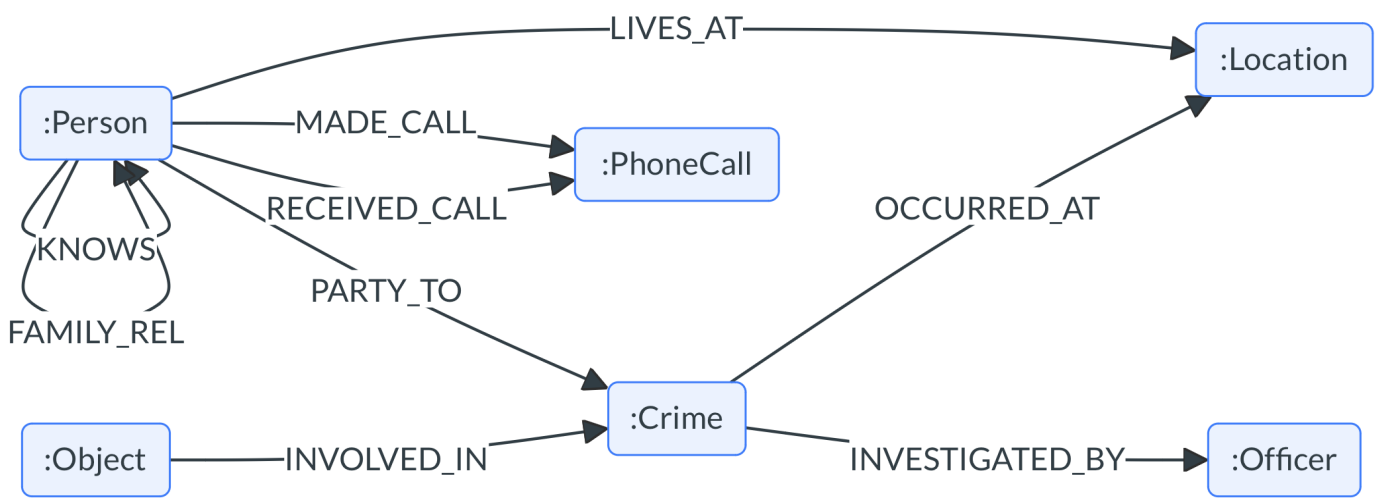

Pole KG schema

**Usage**

In BioChatter Next, we first activate the KG functionality by clicking on the `KG Settings` button in the sidebar. In the settings dialog, we can activate the KG functionality and select how many results we want to retrieve.

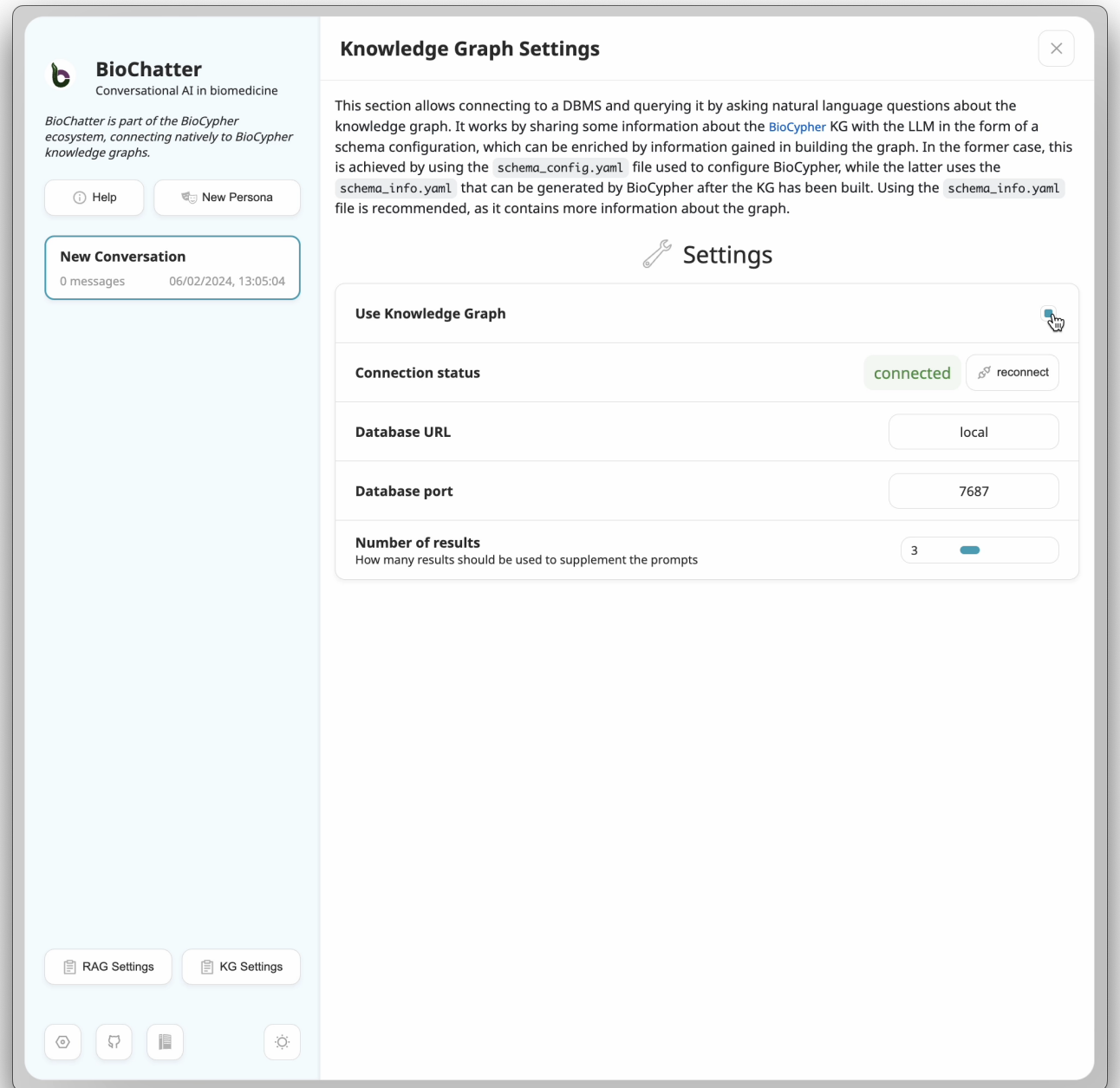

KG Settings

Returning to the conversation and enabling the KG functionality for the current chat (directly above the send button), we can then ask the model about the KG. The language model we use is `gpt-3.5-turbo`. The full conversation is pasted below, including the queries generated by BioChatter.

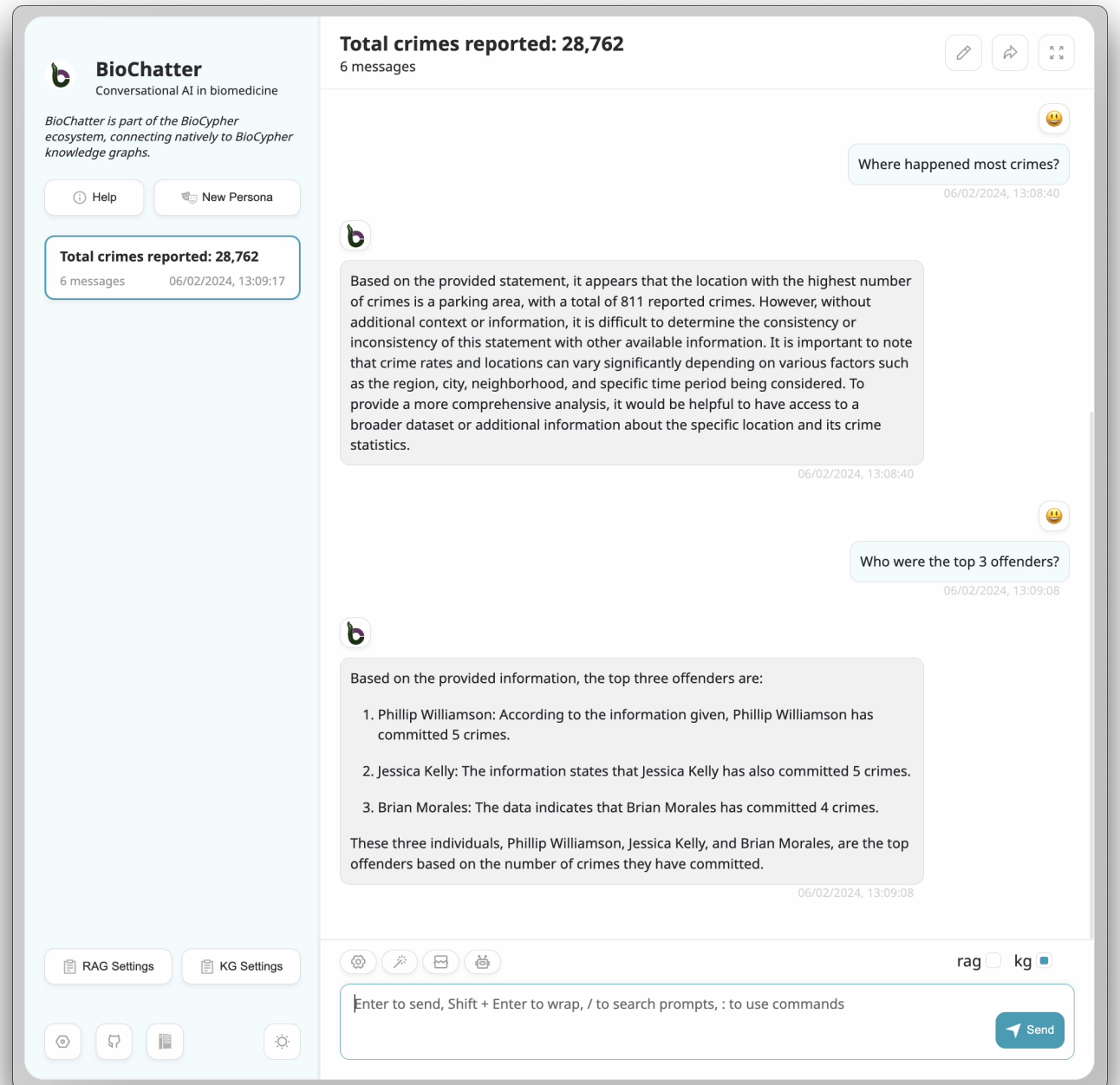

KG Conversation

In the background, the RagAgent module of BioChatter receives the question and generates a query to retrieve the desired information. This is then passed back to the primary model, which includes it in its answer generation.

## Conclusion

The native integration of BioCypher KGs into the BioChatter framework allows for a seamless integration of KGs into the conversational AI. This in turn facilitates knowledge accessibility in a wide range of application domains.

**Note: the apparent inability of GPT to understand certain directionalities, and how BioChatter compensates for this**

Interestingly, while `gpt-3.5-turbo` mostly does a formidable job at translating natural language questions into Cypher queries, it is remarkably obtuse in certain instances. For instance, for the relationship `INVESTIGATED_BY`, which connects a `Crime` to an `Officer`, GPT consistently fails to understand that the relationship implies that the `Officer` is the one who investigates the `Crime`. Instead, it consistently interprets the relationship as if the `Crime` investigates the `Officer`: it consistently proposes the query `MATCH (o:Officer)-[:INVESTIGATED_BY]->(c:Crime) RETURN c, o` instead of the correct `MATCH (c:Crime)-[:INVESTIGATED_BY]->(o:Officer) RETURN c, o`. We were not able to change this behaviour with any contextual prompt instructions.

For this reason, the BioChatter `prompts.py` module uses the knowledge we have about the directionality of edges in the BioCypher KG to only propose options for patterns that actually exist in the KG. In the instance of `INVESTIGATED_BY`, this is the corresponding YAML definition in BioCypher:

```yaml
investigated by:
    is_a: [fact, core]
    represented_as: edge
    label_as_edge: INVESTIGATED_BY
    input_label: INVESTIGATED_BY
    source: crime
    target: officer
```

The presence of the `source` and `target` annotations allows us to provide only the correct options to the LLM, which in turn allows the LLM to generate the correct query.

Strikingly, this even applies to `GPT-4` and `ChatGPT`, which we used to generate the Schema diagram above. The model immediately proposed a correct Mermaid diagram for most of the entire schema definition YAML file, except for the relationship between `Crime` and `Officer`, which it reversed. Via the YAML we provide in the prompt, GPT has access to the correct directionality, but it still fails to understand it.

```
graph LR;
    Person(:Person) -- KNOWS --> Person
    Person -- FAMILY_REL --> Person
    Person -- LIVES_AT --> Location(:Location)
    Person -- PARTY_TO --> Crime(:Crime)
    Person -- MADE_CALL --> PhoneCall(:PhoneCall)
    Person -- RECEIVED_CALL --> PhoneCall
    Officer(:Officer) -- INVESTIGATED_BY --> Crime
    Crime -- OCCURRED_AT --> Location
    Object(:Object) -- INVOLVED_IN --> Crime

    classDef node fill:#f9f,stroke:#333,stroke-width:2px;
    class Person,Officer,Location,Crime,PhoneCall,Object node;
```

Here is the [full chat](#) for reference.

## Conversation

The boxes below the BioChatter responses in the conversation show the queries generated by the RagAgent module.

## Message From You

How many crimes happened in total?

## Message From BioChatter

According to the provided statement, the total number of crimes is 28,762. This information suggests that there have been 28,762 reported crimes in the given context. However, without additional information or context, it is difficult to determine the specific time period, location, or nature of these crimes. It's important to note that crime statistics can vary depending on the source, methodology, and reporting practices. Therefore, it is always advisable to consult official crime reports or local law enforcement agencies for the most accurate and up-to-date information on crime statistics.

```
MATCH (c:Crime)
RETURN count(c) AS numberOfCrimes
```

## Message From You

Where happened most crimes?

## Message From BioChatter

Based on the provided statement, it appears that the location with the highest number of crimes is a parking area, with a total of 811 reported crimes. However, without additional context or information, it is difficult to determine the consistency or inconsistency of this statement with other available information. It is important to note that crime rates and locations can vary significantly depending on various factors such as the region, city, neighborhood, and specific time period being considered. To provide a more comprehensive analysis, it would be helpful to have access to a broader dataset or additional information about the specific location and its crime statistics.

```
MATCH (c:Crime)-[:OCCURRED_AT]->(l:Location)
RETURN l.address AS location, COUNT(c) AS crimeCount
ORDER BY crimeCount DESC
```

## Message From You

Who were the top 3 offenders?

## Message From BioChatter

Based on the provided information, the top three offenders are:

1. Phillip Williamson: According to the information given, Phillip Williamson has committed 5 crimes.

2. Jessica Kelly: The information states that Jessica Kelly has also committed 5 crimes.

3. Brian Morales: The data indicates that Brian Morales has committed 4 crimes.

These three individuals, Phillip Williamson, Jessica Kelly, and Brian Morales, are the top offenders based on the number of crimes they have committed.

```
MATCH (p:Person)-[:PARTY_TO]->(c:Crime)
RETURN p.name, p.surname, COUNT(c) AS num_crimes
ORDER BY num_crimes DESC
LIMIT 3
```

## Supplementary Note 2: Retrieval-Augmented Generation

This vignette demonstrates the RAG module of BioChatter as used by the BioChatter Next application. This basic use case involves an LLM manuscript reading assistant and a vector database with embedded scientific manuscripts [48,49,50,51]. The manuscripts are recent reviews on the topic of ABC transporters (active transport proteins that often serve as efflux pumps in many tissues, with a major function of clearing exogenous substances) in drug resistance phenomena, which we currently investigate in our DECIDER cohort. Manuscripts such as these can be absent from the knowledge base of LLMs for various reasons, such as their recency or the fact that they are not open access. RAG opens up the possibility to retrieve relevant information from these manuscripts, and to inject it into the LLM's generation process.

### Usage

In BioChatter Next, we first activate the RAG functionality by clicking on the `RAG Settings` button in the sidebar. In the settings dialog, we can activate the functionality and upload an arbitrary number of documents, which is only limited by the scale of the vector database system. In this case, and for demonstration purposes, we uploaded the four manuscripts, which leads to the state shown in the screenshot below.

RAG Settings

Note that we chose to split the documents into fragments of 1000 characters each. We could have split by tokens instead, or changed the fragment length and overlap. Fragment length, overlap, and splitting by tokens or characters are decisions that should be made in tandem with the choice of LLM model, and with the number of fragments that should be retrieved for each query. Most importantly, the total text length should not exceed the input context length of the model. Here, we choose to inject 10 fragments per query.

We can now start a new conversation (using the `New Persona` button in the sidebar), for which we select the `Research manuscript helper (RAG)` persona, which includes suitable contextual prompts. Upon activating the RAG in the chat (to signal that we wish to perform RAG for our question), we can enter into a dialogue with the assistant. We use `gpt-3.5-turbo` in this example, which is a very affordable model.

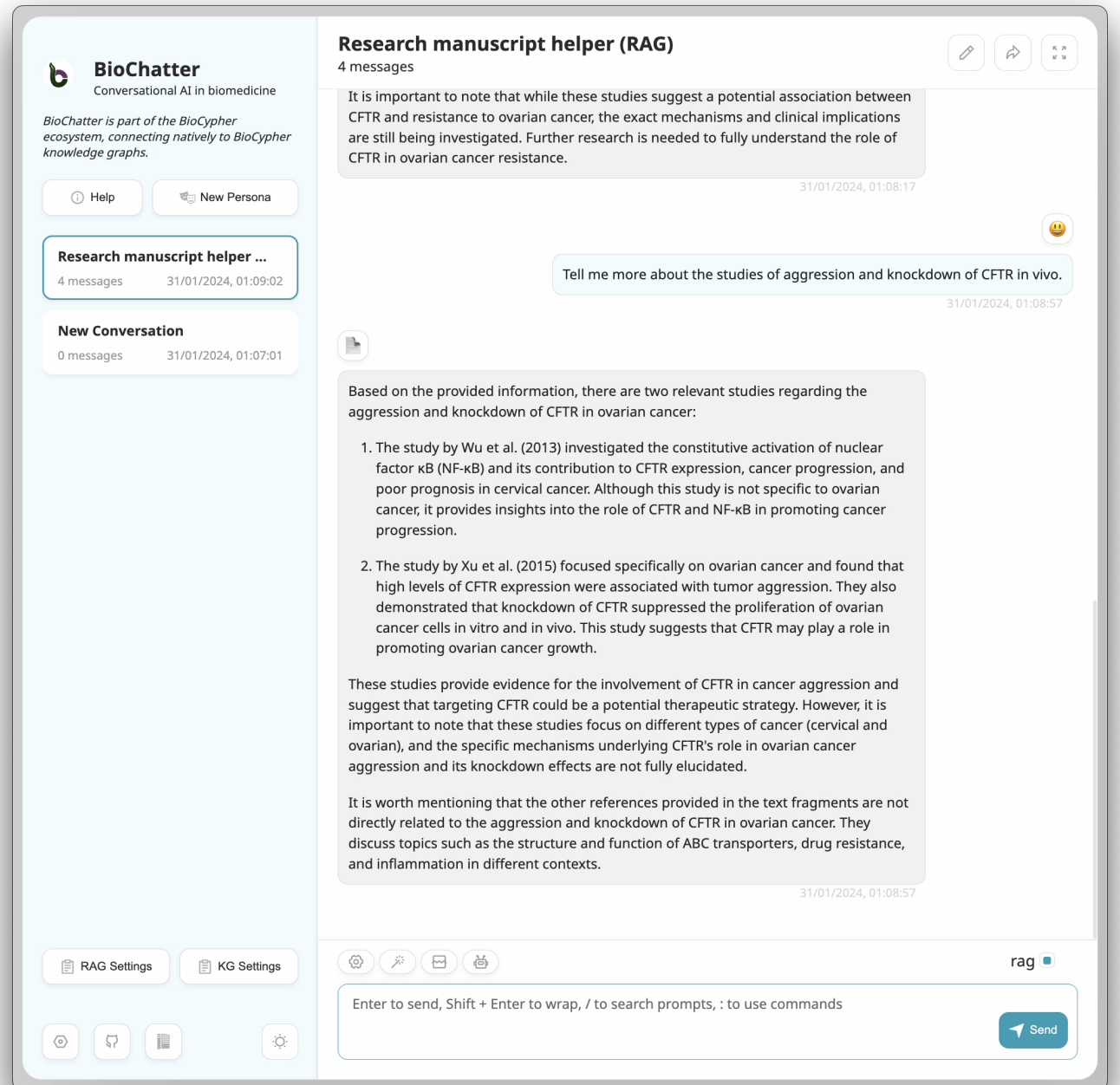

RAG Conversation

## Comparison with ChatGPT

We can ask ChatGPT 4 the same question (only subscription access). By employing web search, ChatGPT 4 is able to find the same study that was the RAG result we were asked about in our follow-up question (Xu et al. 2015); however, the answer is not as detailed as the answer provided by the RAG process, and includes fewer mentions of explicit studies. While it includes sensible information, there is no way of validating most of the claims made by the model. You can read the full answer [here](). Overall, the quality of ChatGPT 4 is comparable to the RAG output, but the user has less control over the knowledge used, and the answer cannot include information that is not easily found on the web (for instance due to access restrictions). Of note, we used `gpt-3.5-turbo` in this example, which is less recent and much less expensive than GPT-4.

Asking the naive ChatGPT 3.5 (free access), the same question yields a very different answer. The free version of ChatGPT cannot search the internet, and does not provide any citations for the general answer it provides. You can read the full answer [here](). The three full conversations are copied below.

## Conclusion

In conclusion, the RAG procedure we implement is able to dramatically improve the quality of GPT-3 answers, making them comparable to GPT-4 with web search, and even surpassing it in some aspects. We invite you to try it out yourself, as we provide an instance of BioChatter Next, including GPT access and a vector database instance, free of charge at https://next.biochatter.org.

## Conversations

### BioChatter RAG

**Contextual Prompt**

You are an assistant to a biomedical researcher.

**Contextual Prompt**

You have access to text fragments from one or multiple scientific manuscripts, which were retrieved by their semantic similarity to the user's question.

**Contextual Prompt**

Your task is to help the user understand the manuscripts and their relevance to their questions. Be concise in your explanations, provide quotes from the manuscripts where appropriate, and be open about limitations in cases where the text does not contain pertinent information.

**Message From You**

What is known about the CFTR protein in resistance to ovarian cancer?

**Message From BioChatter**

Based on the provided information, there are several studies that suggest a potential role of the CFTR (Cystic Fibrosis Transmembrane Conductance Regulator) protein in resistance to ovarian cancer. Here are some key findings from the referenced manuscripts:

1. Gene expression profiling of ovarian tumors before and after chemotherapy revealed molecular signatures of chemoresistant tumors (M, et al., 2006). This study did not specifically mention CFTR, but it provides insights into the molecular changes associated with chemoresistance.

2. ABCF2 expression, which is related to CFTR, was investigated in clear cell adenocarcinoma of the ovary. The study found a relationship between ABCF2 expression and response to chemotherapy or prognosis (Tsuda H, et al., 2010).

3. Alterations in gene copy numbers, including CFTR, were identified as a mechanism for acquired drug resistance in cancer (Yasui K, et al., 2004).

4. ABCF2, an Nrf2 target gene related to CFTR, was found to contribute to cisplatin resistance in ovarian cancer cells (Bao L, et al., 2017).

5. CFTR was not specifically mentioned in the context of ovarian cancer, but a study on hepatocellular carcinoma found that silencing circ-TCF4.85 inhibited cancer progression through microRNA-486-5p-targeted inhibition of ABCF2 (Gao J, et al., 2020).

6. Constitutive activation of nuclear factor κB (NF-κB) was associated with CFTR expression and promoted cancer progression and poor prognosis in cervical cancer (Wu Z, et al., 2013).

7. High levels of CFTR expression were associated with tumor aggression in ovarian cancer, and knockdown of CFTR suppressed ovarian cancer cell proliferation (Xu J, et al., 2015).

It is important to note that while these studies suggest a potential association between CFTR and resistance to ovarian cancer, the exact mechanisms and clinical implications are still being investigated. Further research is needed to fully understand the role of CFTR in ovarian cancer resistance.

**Message From You**

Tell me more about the studies of aggression and knockdown of CFTR in vivo.

**Message From BioChatter**

Based on the provided information, there are two relevant studies regarding the aggression and knockdown of CFTR in ovarian cancer:

1. The study by Wu et al. (2013) investigated the constitutive activation of nuclear factor κB (NF-κB) and its contribution to CFTR expression, cancer progression, and poor prognosis in cervical cancer. Although this study is not specific to ovarian cancer, it provides insights into the role of CFTR and NF-κB in promoting cancer progression.

2. The study by Xu et al. (2015) focused specifically on ovarian cancer and found that high levels of CFTR expression were associated with tumor aggression. They also demonstrated that knockdown of CFTR suppressed the proliferation of ovarian cancer cells in vitro and in vivo. This study suggests that CFTR may play a role in promoting ovarian cancer growth.

These studies provide evidence for the involvement of CFTR in cancer aggression and suggest that targeting CFTR could be a potential therapeutic strategy. However, it is important to note that these studies focus on different types of cancer (cervical and ovarian), and the specific mechanisms underlying CFTR's role in ovarian cancer aggression and its knockdown effects are not fully elucidated.

It is worth mentioning that the other references provided in the text fragments are not directly related to the aggression and knockdown of CFTR in ovarian cancer. They discuss topics such as the structure and function of ABC transporters, drug resistance, and inflammation in different contexts.

## ChatGPT 4

**Anonymous**

What is known about the CFTR protein in resistance to ovarian cancer?

**ChatGPT**

The CFTR protein, known as the cystic fibrosis transmembrane conductance regulator, plays a significant role in ovarian cancer. Research has shown that CFTR is expressed more in ovarian cancer tissues compared to benign ovarian tumors and normal ovaries. This overexpression of CFTR is associated with more advanced stages of ovarian cancer, poor histopathological grade, and higher serum Ca-125 levels, which are all markers of more aggressive disease. Interestingly, the serous type of ovarian cancer shows stronger CFTR staining compared to other types.

Moreover, the role of CFTR extends beyond its traditional function as an ion channel. It interacts with various intracellular signaling pathways that are crucial in cancer progression. For example, in ovarian cancer, CFTR overexpression is linked to interactions with the c-Src signaling pathway, which is involved in cellular growth.

The implications of CFTR expression in ovarian cancer are significant. In studies where CFTR was downregulated, there was a decrease in cell invasion, migration, proliferation, adhesion, and colony formation in vitro. In vivo experiments also showed that the tumorigenic abilities of ovarian cancer cells were significantly repressed when CFTR expression was reduced.

These findings highlight the complex role of CFTR in ovarian cancer and suggest that it could be a potential target for therapeutic intervention. However, it is important to note that the role of CFTR in cancer is multifaceted and varies among different types of cancers. Therefore, a detailed understanding of its function in specific cancer types, like ovarian cancer, is crucial for developing targeted treatments (https://pubmed.ncbi.nlm.nih.gov/25738998/, https://jmhg.springeropen.com/articles/10.1186/s43042-022-00308-7).

## ChatGPT 3.5

**Anonymous**

What is known about the CFTR protein in resistance to ovarian cancer?

**ChatGPT**

As of my last knowledge update in January 2022, there is limited direct information about the CFTR (Cystic Fibrosis Transmembrane Conductance Regulator) protein's role in resistance to ovarian cancer. CFTR is primarily associated with cystic fibrosis, a genetic disorder that affects the lungs and other organs. However, some research suggests potential links between CFTR and cancer, including ovarian cancer, although the mechanisms are not fully understood.

Here are some key points to consider:

CFTR Expression: Studies have shown that CFTR is expressed in various tissues, including the female reproductive system. Abnormal CFTR expression or function could potentially impact the development or progression of ovarian cancer. Research is ongoing to understand the role of CFTR in cancer biology. CFTR Mutations: Certain mutations in the CFTR gene have been associated with an increased risk of cancer, although the evidence for ovarian cancer specifically is limited. Mutations in CFTR are more commonly known for their role in cystic fibrosis. Individuals with cystic fibrosis have an increased risk of various cancers, including gastrointestinal cancers, but the link to ovarian cancer is not well-established. Ion Transport and Cancer: CFTR is involved in ion transport across cell membranes, and this function could potentially influence cellular processes relevant to cancer. Changes in ion transport can affect cell proliferation, apoptosis, and other cellular functions that are important in cancer development and progression. Potential Therapeutic Target: Some research suggests that modulating CFTR activity or expression could be a potential target for cancer therapy, including ovarian cancer. However, more research is needed to determine the feasibility and effectiveness of such approaches. It's important to note that the understanding of the role of CFTR in cancer, including ovarian cancer, is an evolving field, and new research findings may have emerged since my last update in January 2022. To get the most up-to-date information on this topic, I recommend consulting recent scientific literature and speaking with medical experts or researchers actively working in the field of ovarian cancer and CFTR biology.